\documentclass[preprint2]{aastex62}

\usepackage{mathtools}

\graphicspath{{./}{figures/}}
\usepackage{lineno}

\shorttitle{Multipolar EM emission of Magnetar}
\shortauthors{Wang et al.}

\begin{document}

\title{Multipolar Electromagnetic Emission of Newborn Magnetar}

\correspondingauthor{ Rahim Moradi, Yu Wang,\\ Liang Li\\}

\author{Yu Wang}
\email{rmoradi@ihep.ac.cn}
\affil{ICRA and Dipartimento di Fisica, Sapienza Universit\`a di Roma, P.le Aldo Moro 5, 00185 Rome, Italy.}
\affil{ICRANet, P.zza della Repubblica 10, 65122 Pescara, Italy.}
\affil{INAF -- Osservatorio Astronomico d'Abruzzo, Via M. Maggini snc, I-64100, Teramo, Italy.}

\author{Rahim Moradi}
\email{yu.wang@inaf.it}
\affil{Key Laboratory of Particle Astrophysics, Institute of High Energy Physics, Chinese Academy of Sciences, Beijing 100049, China.}

\author{Liang Li}
\email{liang.li@icranet.org}
\affil{ICRA and Dipartimento di Fisica, Sapienza Universit\`a di Roma, P.le Aldo Moro 5, 00185 Rome, Italy.}
\affil{ICRANet, P.zza della Repubblica 10, 65122 Pescara, Italy.}
\affil{INAF -- Osservatorio Astronomico d'Abruzzo, Via M. Maggini snc, I-64100, Teramo, Italy.}

\begin{abstract}

It is generally recognized that the electromagnetic multipolar emission from magnetars can be used to explain radiation from Soft Gamma Repeaters (SGRs) or Anomalous X-ray Pulsars (AXPs), but they have little impact on the spindown of magnetars.  We here present an analytical solution for the neutron star multipolar electromagnetic fields and their associated expected luminosities. We find that for newborn millisecond magnetars, the spin-down luminosity from higher multipolar components can match or even exceed that from the dipole component. Such high-intensity radiation will undoubtedly affect related astrophysical phenomena at the birth of a magnetar. We show that the spin-down luminosity from multipoles can well explain the majority of Gamma-Ray Bursts (GRBs) afterglows, from the plateau starting at several hundred seconds till the normal decay phase lasting for many years. The fitted magnetar parameters for GRB afterglows are all typical values, with spins in the millisecond range and magnetic field strengths in the order of $10^{14} - 10^{15}$ Gauss. Our results in turn, provide support for the hypothesis that GRBs originate from the birth of magnetars with a few millisecond period, thus deepening our understanding of the complex magnetic field structure and the equation of state of magnetars.

\end{abstract}

%\keywords{editorials, notices --- miscellaneous --- catalogs --- surveys}

\section{Introduction} 
\label{sec:intro}

Our understanding of neutron stars (NSs) is largely derived from phenomena related to {their} magnetic fields \citep{2004Sci...304..536L, 2007PhR...442..109L,2010PNAS..107.7147K,2021Univ....7..351I}. Since, apart from the rare gravitational wave observations 
 \citep{2017ApJ...848L..13A}, the low-frequency radio and high-energy X-ray pulses, as well as the higher-energy transitional flares, are all related to their strong magnetic fields, which generate electromagnetic radiation either by accelerating charged particles or through its own dissipation \citep{1992herm.book.....M,2005puas.book.....L,2021ASSL..461...97E}.

Research on the magnetic fields of NSs has historically focused on the dipole component for two primary reasons{.} First, the dipole model has a simple mathematical form, of which the first-order approximation sufficiently accounts for the majority of observational data \citep{1969Natur.221..454G, 1982RvMP...54....1M}. Indeed, the inaugural observation of the NS in 1968 could be explained by magnetic dipole radiation \citep{1968Natur.217..709H, 1969ApJ...157.1395O}. Second, the strength of higher-order multipoles decreases more rapidly than that of the dipole with distance from the NS. Consequently, the dipole normally dominates the radiation at large distances. Most of our observations of NSs, such as the spin-down rate and the pulsar wind geometry, are of far-field radiation, which can be effectively explained by dipole radiation. Detecting the signatures of higher-order multipoles emanating from small-size regions which require high-resolution observations is challenging \citep{1993Ap&SS.200..251K,2022ARA&A..60..495P}.

Thanks to the refinement of theoretical models, increased complexity of computer simulations, and continuous development of observational techniques, studies are increasingly integrating multipolar components to explain and predict the complex behaviors observed in NSs, particularly in strongly magnetized NS{s} known as magnetars \citep{2015RPPh...78k6901T,2017ARA&A..55..261K,2020NewAR..9101544P,2021Univ....7..351I,2022ARA&A..60..495P,2024Galax..12....6T}.

Theoretically, multipole fields near the surface are generally required to facilitate the production of electron-positron pairs in the magnetosphere \citep{1975ApJ...196...51R}. And the theory that the polar cap is anchored with a {multipolar} field \citep{2006ApJ...650.1048G} is supported by the discovery that the area heated by X-ray{s} is significantly smaller than traditional polar cap size \citep{2005ApJ...624L.109Z}. Observationally, SGR 0418+5729 presents a typical case. Its complex and strong multipolar fields, of strength as high as $10^{15}$ Gauss measured by the proton cyclotron lines \citep{2013Natur.500..312T}, produce the observed X-ray bursts \citep{2010Sci...330..944R, 2011ApJ...740..105T}. While its spindown reveal a dipolar magnetic field of only $B_p \simeq 6 \times 10^{12}$ G \citep{2013ApJ...770...65R}. The multipole components can also channel energy and heat to localized areas on the surface of NSs, creating hot spots \citep{2014ApJ...784..168H, 2014MNRAS.444.3198G, 2019ApJ...887L..21R,2020ApJ...889..165D,2021ApJ...907...63K, 2021ApJ...914..118D}. The detection of thermal X-ray pulsations from the hot surface of PSR J0030+0451 by the NS Interior Composition Explorer (NICER) telescope suggests that pulsars are originated from the global-scale multipolar magnetic fields \citep{2019ApJ...887L..23B}. In addition, {multipoles have been proposed to explain} the observations of flare activities, irregular timing properties, detailed X-ray analysis, and etc \citep[see e.g.][]{2014ApJ...786...62S,2015RPPh...78k6901T,2021NatAs...5..145I,2021MNRAS.507.2208W,2023MNRAS.523.4089S,2024arXiv240307649F}.

The examples provided above primarily focus on multipolar radiation characteristics occurring at least a century after the formation of the NS. However, to comprehensively understand the formation of magnetars, the complex magnetic structure, the equation of state, and their evolution, observations from hundreds of years later are insufficient. Early-stage information is needed to gain a more complete understanding. But as of now, early information is not sufficiently discovered. The Chandra Deep Field-South survey (CDF-S) has provided the sole recognized early-stage observation, capturing the X-ray transient known as CDF-S XT2, a magnetar just born from the binary NS merger \citep{2019Natur.568..198X}.

To obtain early information about magnetars, our approach is to examine multipolar radiation, particularly starting from the moment of their formation. Magnetars can theoretically form from binary NS mergers and from single star collapse. The formation of magnetar{s} is associated with and plays a critical role in various areas of astrophysics, including {superluminous supernovae (SLNSe)}, gamma-ray bursts (GRBs) and fast radio bursts (FRBs).

SLSNe are supernovae that are approximately 100 times more luminous than typical supernovae. Some SLSNe are believed to be powered by newly formed magnetars, with the rotational energy being converted into electromagnetic radiation, enhancing the brightness of the supernova explosion \citep{2010ApJ...719L.204W,2010ApJ...717..245K}. FRBs are brief, intense extragalactic radio flashes, and one hypothesis suggests they could be related to magnetars, either through the magnetars' own activities or processes associated with their formation \citep{2022A&ARv..30....2P,2023RvMP...95c5005Z}. GRBs are intense flashes of gamma-rays that can last from a fraction of a second to several minutes, followed by an X-ray afterglow that persists for years \citep{2004RvMP...76.1143P,2014ARA&A..52...43B}. In this paper we will specifically focus on GRBs to highlight the significance of newborn magnetars in astrophysical events.

This article is structured to flow logically, beginning with Section \ref{sec:model} and \ref{sec:evolution}, where we discuss magnetar emissions: we examine the theoretical model of multipolar emission from a newborn magnetar and present the expected luminosity lightcurve by demonstrating some cases. Following this, Section \ref{sec:grb} offers a concise overview of phenomena related to magnetars, with a special focus on GRB{s}. In Section \ref{sec:application}, we apply the magnetar multipolar model to the afterglow observations of GRBs, uncovering a consistency between the predicted magnetar evolution and the observed plateau and subsequent normal decay phases of GRB afterglows. Finally, Section \ref{sec:conclusion} delves into our findings, examining their limitations and broader implications.

%%%%%%%%%%%%%%%%%%%%%%%%%%%%%%%%%%%%%%%%%%%%%%%%
\section{Modeling Multipolar Spin-down and Emission}\label{sec:model}
%%%%%%%%%%%%%%%%%%%%%%%%%%%%%%%%%%%%%%%%%%%%%%%%

The electromagnetic field of NS{s} is described by the Maxwell equations, the equations can be decomposed then solved by introducing a set of vector spherical harmonics \citep{1998clel.book.....J,barrera1985}. Each harmonic mode is identified by a set of multipole order number{s} $l$ and azimuthal mode number{s} $m$. Therefore, we have the magnetic multipoles, including dipole ($l=1$), quadrupole ($l=2$), hexapole ($l=3$), octopole ($l=4$) and higher orders. Considering a rotating NS in {the} vacuum, \citet{ptri2015} gives a complete analytical solution of the NS multipolar electromagnetic fields, which is adopted in this article. For simplicity, instead of considering the finite size of the NS, we take the simplified solution treating NS as a point-like source. 

\vspace{0.5cm}
\noindent \textbf{Magnetic dipole} is the lowest multipolar component, which has been  well studied and is considered as the only magnetic field in the majority GRB articles concerning the energy injection from a NS or a magnetar \citep[for e.g.][]{Dai1998,2001ApJ...552L..35Z,2015RPPh...78k6901T,2018ApJS..236...26L,2007ApJ...670..565L}. The spin-down energy loss rate (luminosity) by Poynting flux is
\begin{equation}
    L_{\rm dip} = \frac{2}{3 \,c^3} \, \Omega^4\,B_{\rm dip}^2\,R^6 \, \Theta^2_{\rm dip}
\end{equation}
\begin{equation}
    \Theta^2_{\rm dip} = \sin^2\chi_1
\end{equation}
where $\Omega$ is the angular velocity of the NS, $B_{\rm dip}$ is the dipole magnetic field strength, $R$ is the radius of the NS, $c$ is the speed of light, $\Theta_{\rm dip}$ contains the angle-related terms, $\chi_1$ is the inclination {angle} of the magnetic dipole moment, related to the azimuthal mode number $m$, $\chi_1 = 0$ or $90$ degrees for $m=0$ or $1$. The rotational energy of the NS $E_{\rm rot}$ {evolves} as
\begin{equation}
    \frac{dE_{\rm rot}}{dt} = I \Omega \dot{\Omega } = - L
\label{eq:dE_dt}
\end{equation}
where $I$ is the {moment} of inertia and $L$ is the energy loss rate. For a pure dipole $L = L_{\rm dip}$, the above equation gives the solution of the 
evolution of the angular velocity
\begin{equation}
    \Omega = \Omega_0 (1+\frac{t}{\tau_{\rm dip}})^{-1/2},
\end{equation}
and the energy loss rate
\begin{equation}
    L_{\rm dip} = L_{\rm dip,0} (1+\frac{t}{\tau_{\rm dip}})^{-2}
\end{equation}
where $\Omega_0$ is initial angular velocity, $\tau_{\rm dip}$ characterizes the time scale of the NS losing half of its rotational energy
\begin{equation}
    \tau_{\rm dip} = \frac{3 I c^3}{4 \Omega_0^2 B_{\rm dip}^2 R^6 \Theta^2_{\rm dip}}
\end{equation}
and $L_{\rm dip,0}$ is the initial energy loss rate
\begin{equation}
    L_{\rm dip,0} = \frac{I \Omega_0^2}{2 \tau_{\rm dip}}
\end{equation}

The angular velocity $\Omega$ decreases slowly before $\tau_{\rm dip}$, then follows almost a power-law decay with index $-1/2$. Correspondingly, the energy loss rate $L_{\rm dip}$ keeps almost a constant before $\tau_{\rm dip}$, then drops with power-law index $-2$, since the dipole energy loss $L_{\rm dip} \propto \Omega^4$. The characteristic timescale $\tau_{\rm dip}$ relates to all the properties of the NS, a more massive, slower spin, lower magnetic field, smaller radius NS corresponds to a longer $\tau_{\rm dip}$. The initial energy loss rate $L_{\rm dip,0}$ is simply in the form of the total rotational energy $I \Omega_0^2/2$ divided by the characteristic timescale $\tau_{\rm dip}$.

\vspace{0.5cm}
\noindent \textbf{Magnetic quadrupole} can be easily generated, for example, {as the} magnetic dipole moment of a neutron star is not aligned with its rotation axis. Given a quadrupole magnetic field $B_{\rm quad}$, the energy loss rate is 
\begin{equation}
    L_{\rm quad} = \frac{32}{135\,c^5} \, \Omega^6 \, B_{\rm quad}^2 \, R^8 \, \Theta^2_{\rm quad},
\end{equation}
\begin{equation}
   \Theta^2_{\rm quad} = \sin^2\chi_1 \, ( \cos^2\chi_2 + 10 \, \sin^2\chi_2 )
\end{equation}
where the angular term $\Theta_{\rm quad}$ contains two inclination angles, $\chi_1$ reminisces the dipole field, $\chi_2$ relates to the quadrupole field, the different choice of $\chi_2= 0$ or $90$ degrees brings {a factor $10$ of difference}. For a pure magnetic quadrupole, $L = L_{\rm quad}$, equation \ref{eq:dE_dt} gives the evolution of the spin
\begin{equation}
    \Omega = \Omega_0 (1+\frac{t}{\tau_{\rm quad}})^{-1/4},
\end{equation}
and the loss rate of the rotational energy
\begin{equation}
    L_{\rm quad} = L_{\rm quad,0} (1+\frac{t}{\tau_{\rm quad}})^{-3/2}
\end{equation}
where the characteristic timescale of quadrupole emission
\begin{equation}
    \tau_{\rm quad} = \frac{135 I c^5}{128 \Omega_0^4 B_{\rm quad}^2 R^8 \Theta^2_{\rm quad}}
\end{equation}
and the initial energy loss rate
\begin{equation}
    L_{\rm quad,0} = \frac{I \Omega_0^2}{4 \tau_{\rm quad}}
\end{equation}

Compared to the dipole field, the energy loss rate of quadrupole field is more {sensitive} to the spin of the NS, since $L_{\rm quad} \propto \Omega^6$ and  $L_{\rm dip} \propto \Omega^4$. Consequently, the energy loss rate $L_{\rm quad}$ at time later than the characteristic timescale $\tau_{\rm quad}$ follows a shallower power-law decay than the dipole field, with power-law index $-3/2$. 

\vspace{0.5cm}
\noindent \textbf{Magnetic hexapole, octopole and higher-order {multi}poles} are from high-order harmonic modes. The spin-down luminosities of hexapole and octopole are
\begin{equation}
    L_{\rm hexa}  = \frac{2}{4725\,c^7} \, \Omega^8\,B_{\rm hexa}^2\,R^{10} \Theta^2_{\rm hexa}
\end{equation}
and
\begin{equation}
    L_{\rm octo} = \frac{4}{297675\,c^9} \, \Omega^{10}\,B_{\rm octo}^2\,R^{12} \Theta^2_{\rm octo},
\end{equation}
respectively. The related angular terms are more complicated, for hexapole field,
\begin{eqnarray}
&&\Theta^2_{\rm hexa} = \sin^2\chi_1  ( 29 \, \cos^2\chi_2 +  \sin^2\chi_2 \,\\&& ( 1664 \, \cos^2\chi_3  + 15309\, \sin^2\chi_3 ) ), \nonumber 
\end{eqnarray}
an additoinal inclination angle $\chi_3$ is involved, $\Theta^2_{\rm hexa} = \{29,1664,15309\}$ for the configurations of $\{\chi_1, \chi_2, \chi_3\} = \{(90,0,0), (90,90,0), (90,90,90)\}$ degrees. The octopol{a}r field has one more angle $\chi_4$,
\begin{eqnarray}
  &&\Theta^2_{\rm octo} = \sin^2 \chi_1 \, 
  ( 23 \, \cos^2\chi_2 + \sin^2\chi_2 \, \nonumber \\&& ( 5632 \, \cos^2\chi_3 + \sin^2\chi_3 \, ( 133407 \, \cos^2\chi_4 \\&& +  1179648 \, \sin^2\chi_4 ) ) ), \nonumber
\end{eqnarray}
which gives $\Theta^2_{\rm octo} = \{23,5632,133407,1179648\}$ for the configuration of $\{\chi_1, \chi_2, \chi_3, \chi_4\}$ = $\{(90,0,0,0),$ $ (90,90,0,0),$ $ (90,90,90,0), $ $(90,90,\nonumber\\90,90)\}$ degrees.

For a pure hexapole, the spin and the luminosity change as
\begin{equation}
    \Omega = \Omega_0 (1+\frac{t}{\tau_{\rm hexa}})^{-1/6},
\end{equation}
\begin{equation}
    L_{\rm hexa} = L_{\rm hexa,0} (1+\frac{t}{\tau_{\rm hexa}})^{-4/3}
\end{equation}
where the characteristic timescale and the initial luminosity are
\begin{equation}
    \tau_{\rm hexa} = \frac{4725 ~ I c^7}{12 \Omega_0^6 B_{\rm hexa}^2 R^{10} \Theta^2_{\rm hexa}}
\end{equation}
\begin{equation}
    L_{\rm hexa,0} = \frac{I \Omega_0^2}{6 \tau_{\rm hexa}}
\end{equation}

For the energy loss by a pure octopole, the spin and the luminosity decreases as 
\begin{equation}
    \Omega = \Omega_0 (1+\frac{t}{\tau_{\rm {octo}}})^{-1/8},
\end{equation}
\begin{equation}
    L_{\rm {octo}} = L_{\rm {octo},0} (1+\frac{t}{\tau_{\rm {octo}}})^{-5/4}
\end{equation}
where the characteristic timescale and the initial luminosity are
\begin{equation}
    \tau_{\rm {octo}} = \frac{297675 ~ I c^9}{32 \Omega_0^8 B_{\rm {octo}}^2 R^{12} \Theta^2_{\rm {octo}}}
\end{equation}
\begin{equation}
    L_{\rm {octo},0} = \frac{I \Omega_0^2}{8 \tau_{\rm {octo}}}
\end{equation}

Generally, for multipole with order number $l$, the luminosity is in the form of
\begin{equation}
    L_{l}  =C_{l} \Omega^{2l+2}\,B_{l}^2\,R^{2l+4} \Theta^2_{l}
\label{eq:spin-down-luminosity}
\end{equation}
where $C_{l}$ is {a} constant, $B_{l}$ is the magnetic field, $\Theta_{l}$ is the angular term. The time evolution of the angular velocity and the luminosity are
\begin{equation}
    \Omega = \Omega_0 (1+\frac{t}{\tau_{l}})^{-2l},
\end{equation}
\begin{equation}
    L_{ {l}} = L_{l,0} (1+\frac{t}{\tau_{l}})^{-1-1/l}.
\end{equation}
The characteristic {spin-down} timescale{, which can be considered as the duration of the plateau,}  and initial luminosity are
\begin{equation}
    \tau_{l} = \frac{ I c^{2l+1}}{ (2l+2) C_{l} \Omega_0^{2l} B_{l}^2 R^{2l+4} \Theta^2_{l}}
    \label{eq:ctime}
\end{equation}
\begin{equation}
    L_{l,0} = \frac{I \Omega_0^2}{2l \tau_{l}}
\end{equation}

{\section{Magnetic Field Evolution}\label{sec:evolution}}

The magnetic field structure of magnetar is an issue that is continuously being explored \citep{2021Univ....7..351I}. Based on current theories and observations, it is generally recognized that the magnetic field structure is very complex. Besides the large-scale dipolar poloidal field, there can also exist other scales of multipolar poloidal fields and toroidal fields. 

Toroidal fields influence the spin-down rate by affecting the internal dynamics of the star and its interaction with the surrounding environment. This is an indirect and extremely complex process \citep{2018ApJ...852...21G}. Therefore, this article, aiming to provide a first-order approximate picture, does not consider the toroidal fields. 

Regarding the relationship between the dipole and high-order multipoles, it lacks a consensus. First, the formation of magnetic field of magnetars remains a topic of extensive discussion. There are numerous processes proposed to augment and reconfigure the magnetic field during the proto-NS phase \citep{1992ApJ...395..250G, 2002ApJ...574..332T, 2021Univ....7..351I}. Taking the classical dynamo model as an example, there could be an orderly dynamo that generates a strong dipole field, or it could be a stochastic dynamo, which generates strong multipolar fields with a weak dipolar field \citep{1993ApJ...408..194T,2016PNAS..113.3944G,2021Univ....7..351I}. Secondly, the {inferred} magnetic field strengths of magnetars exhibit a wide range. Most magnetars\footnote{McGill Online Magnetar Catalog: \url{http://www.physics.mcgill.ca/~pulsar/magnetar/main.html}} \citep{2014ApJS..212....6O}, identified by SGRs or AXPs, have their dipolar magnetic field strengths calculated to exceed the quantum critical field strength, $B_Q \approx 4.4 \times 10^{13}$ Gauss. These calculations are based on the observed spin periods and spin-down rates, employing the dipole model.  However, there are magnetars with dipolar field below the quantum critical field $B_Q$, such as SGR 0418+5729 \citep{2010Sci...330..944R}, Swift J1822.3-1606 \citep{2012ApJ...754...27R}, and 3XMM J185246.6+003317 \citep{2014ApJ...781L..17R}. Among these, SGR 0418+5729 has {the} lowest dipolar field as $B_p \simeq 6 \times 10^{12}$ Gauss \citep{2013ApJ...770...65R}. The question of whether these lower-strength dipolar magnetic fields are inherently low at the time of formation or have decayed from higher intensities remains under discussion \citep{2011ApJ...732L...4A,2011ApJ...740..105T,2013IJMPD..2230024T}. Theoretical models suggest that the high-order multipoles could be even stronger than the dipole, possibly reaching values up to several times $10^{15}$ Gauss \citep{2016PNAS..113.3944G}. Observationally,  determining the strength of higher-order multipolar is complex, largely because these components mainly influence only the localized phenomena. Their existence and strengths are often inferred from, such as, the distribution of surface hotspots, the details of the emission spectra, and the energetics of flares and outbursts. For instance, although SGR 0418+5729 does not meet the typical magnetar standard in terms of dipole field strength, the observation of cyclotron absorption features implies the existence of a complex magnetic field, with  strengths reaching $10^{15}$ Gauss near the surface \citep{2013Natur.500..312T}. The SGR J1935+2154, associated with FRBs \citep{2021NatAs...5..414K}, has a dipolar magnetic field strength of $4.4 \times 10^{14}$ Gauss, obtained from changes in its spin period, aligning with typical magnetar values. Further analysis of fitting its X-ray spectra with the STEMS3D model, reveals a stronger, local nondipolar magnetic field of $9.6 \times 10^{14}$ Gauss \citep{2020ApJ...905L..31G}. 

{Summarizing, from the observations} magnetars can exhibit both dipole and high-order multipole magnetic field components. The dipole field strength may either exceed or fall below the critical field threshold, while the multipole fields are consistently above it. {For any individual magnetar with observed both dipole and multipole components,} the strength of the multipole field surpasses that of the dipole field. This pattern holds true at least for the currently observed limited number of samples.

These confirmed magnetars are relatively young compared to normal NSs, with ages ranging from a few thousand to tens of thousands of years, magnetars with weaker dipole fields may be older \citep{2015SSRv..191..315M,2021ApJ...913L..12M}. Inferring the initial strength of the magnetic field from observations made thousands of years later is not straightforward, as different theoretical models yield varying results \citep{2021Univ....7..351I}. Generally, magnetic fields decay over time, but it is also possible for initially buried magnetic fields to emerge, leading to an increase in field strength \citep{1999A&A...345..847G}.

In light of the aforementioned considerations, we present three scenarios to visualize the evolution of the quiescent luminosity starting from the birth of the magnetars using the formulae derived in section \ref{sec:model}. We use typical values and do not account for the evolution of the magnetic fields.

\begin{figure}
\centering
\includegraphics[width=1.0\hsize,clip,angle=0]{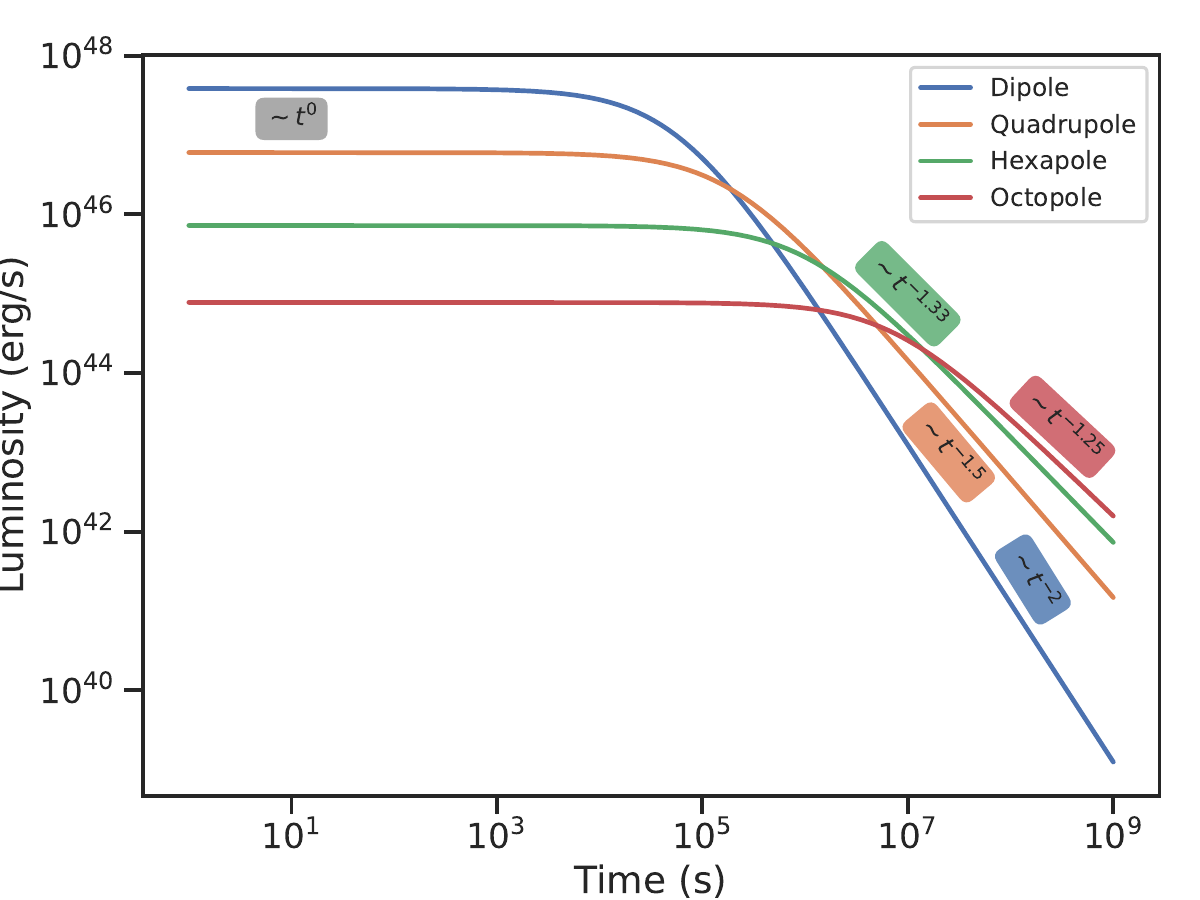}
\caption{The spin-down luminosity of different multipoles with the same magnetic strength ($B = 10^{14}$ G), respectively.  All show a plateau phase, followed by a power-law-like decay, the higher order multipole produces a shallower decay.
}
\label{fig:single-pole}
\end{figure}

The first scenario involves that only one magnetic field component exists. We examine the radiation evolution of dipole, quadrupole, hexapole, and octopole independently. By setting the magnetic field strength of $10^{14}$ Gauss, selecting the maximum values of the angular factors of different multipoles as $\Theta_{\rm dip}^2 = 1, \Theta_{\rm quad}^2 = 10, \Theta_{\rm hexa}^2 = 15338, \Theta_{\rm octo}^2 = 1179671$, along with the initial spin of $1$~ms, the radius of $1.0 \times 10^{6}$~cm and the mass of $1.4 M_\odot$ (the same angular factors, initial spin, radius, and mass values will be applied in subsequent examples), we generate figure \ref{fig:single-pole}.  It can be observed that both dipoles and multipoles feature an initial plateau phase followed by a power-law decay phase. The plateau phase luminosity is highest for the dipole, with the luminosity of multipoles proportionally decreasing as the order increases. Correspondingly, the characteristic age proportionally increases. The power-law index in the decay phase is steepest for the dipole at -2, with the quadrupole, hexapole, and octopole respectively at -1.5, -1.33, and -1.25. {In reality, it is highly unlikely only a single-component magnetic field exists. In particular, there has never been observed a case where only a multipole exists without a dipole. A more realistic scenario is the simultaneous existence of multiple magnetic components.}

It is important to note that most observed magnetars, having evolved over thousands of years, have spin periods on the order of seconds, { and their multipole spindown luminosity is negligible compared to the dipole. However, for newly born magnetars with milliseconds spins shown in figure \ref{fig:single-pole}, the spindown luminosities from adjacent multipolar orders does not differ by orders of magnitudes. This is due to the spindown luminosity of higher-order multipoles is more sensitive to the spin period as $L_{l} \propto \Omega^{2l+2}$ from Eq.~\ref{eq:spin-down-luminosity}}.  Our subsequent examples even demonstrate that under certain magnetic field configurations, multipoles can produce a brighter spindown luminosity than the dipole.

\begin{figure}
\centering
\includegraphics[width=1\hsize,clip,angle=0]{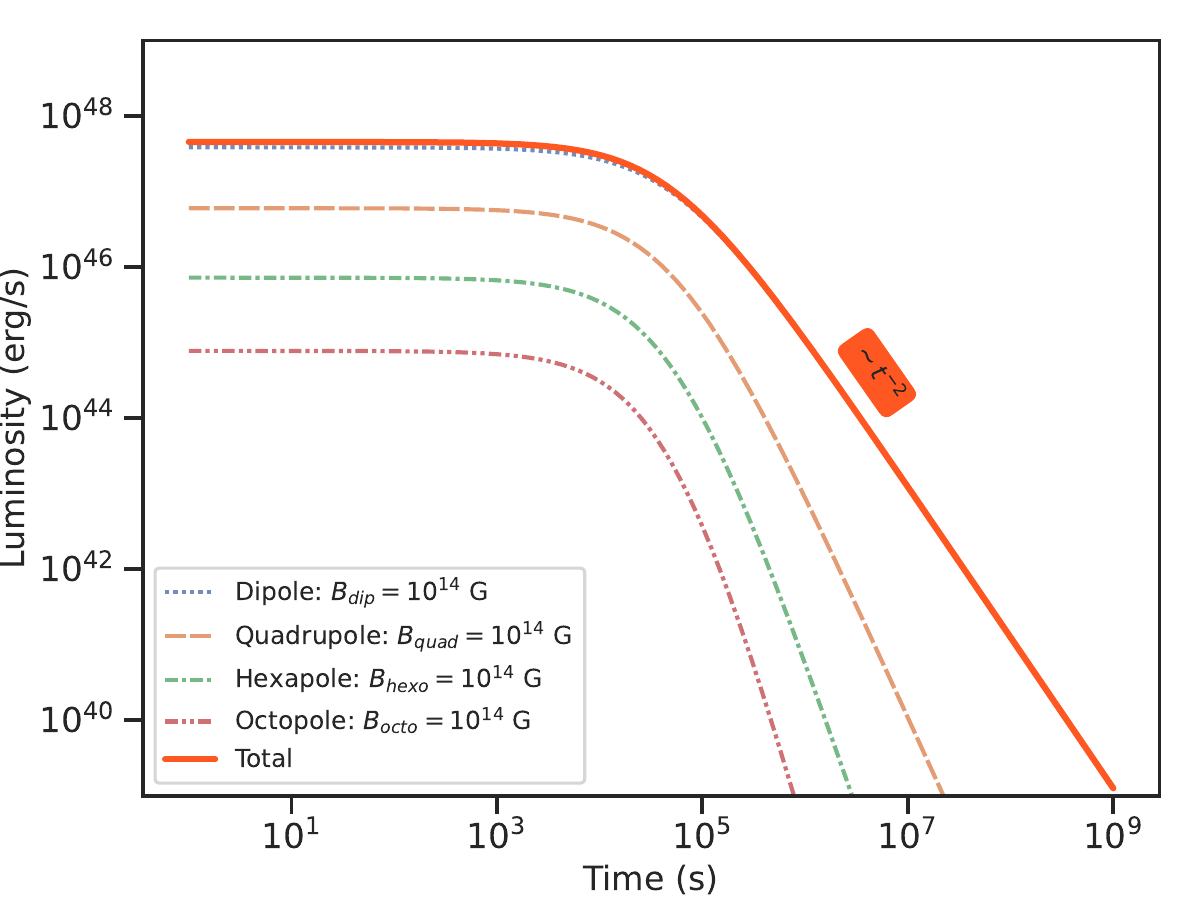}
\caption{A magnetar having dipole, quadrupole, hexapole and octopole, of which the strength of magnetic fields are the same ($B = 10^{14}$ G). The dipole dominates the evolution all the time.}
\label{fig:multipole-example-sameB}
\end{figure}

The second scenario involves the simultaneous existence of the dipole, quadrupole, hexapole, and octopole, with all magnetic field strengths set at $10^{14}$ Gauss. The results, as shown in figure \ref{fig:multipole-example-sameB}, indicate that the dipole dominates the spindown throughout the entire period. The contribution of the multipoles gradually decreases over time, becoming negligible during the decay phase. The most significant difference from figure \ref{fig:single-pole} is that, with all components present, the increase of spin period caused by the dipole leads to earlier arrivals and steeper decays of the decay phase for higher-order multipole components.

\begin{figure}
\centering
\includegraphics[width=1\hsize,clip,angle=0]{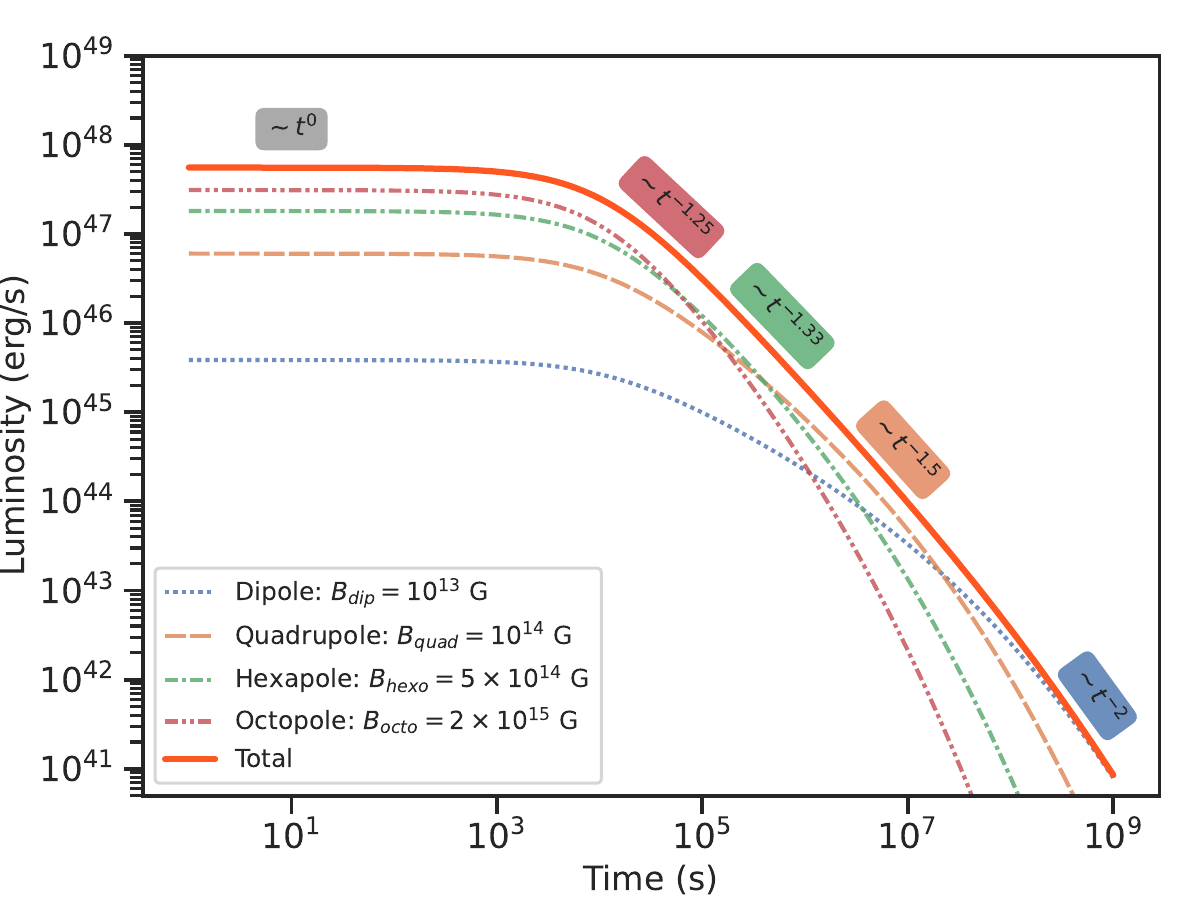}
\caption{A magnetar having dipole, quadrupole, hexapole and octopole. The strength of magnetic field gets higher for the higher order multipole. Initially the high-order multipoles dominate the energy release, after $10^8$ s (about three years), the dipole starts to be dominant.}
\label{fig:multipole-example}
\end{figure}

The third scenario also features the simultaneous presence of the dipole, quadrupole, hexapole, and octopole, but with magnetic field strengths increasing with the order. {This example is to demonstrate a counterintuitive possibility. Generally, for an expansion, it is believed that the higher the order, the smaller it contributes. However, luminosity is not only influenced by the expansion coefficients but also related to the magnetic field strength and the spin period.} We chosen dipole $B_{dip} = 10^{13}$ Gauss, quadrupole $B_{quad} = 10^{14}$ Gauss, hexapole $B_{hex{a}} = 5 \times 10^{14}$ Gauss, octopole $B_{octo} = 2 \times10^{15}$ Gauss {and the initial spin of $1$~ms}. With such a configuration of magnetic fields, we obtain the figure \ref{fig:multipole-example}.  During the plateau phase of this scenario, the higher the order, the higher the spindown luminosity{;} this result differs form the previous scenarios. In the decay phase, as time progresses, the luminosity of the lower-order components gradually surpasses that of the higher order, and the decay index becomes increasingly steep. Eventually, around $10^8$ seconds later, the dipole dominates the spindown. Together with figure \ref{fig:multipole-example-sameB}, we can conclude that under a reasonable magnetic {field} configuration, regardless of the initial strength differences between the dipole and multipole components, the dipole component will eventually dominate the spindown. This means that from the observations of thousands of years after the birth of magnetar, regardless of whether the calculated dipole strength exceeds the critical field strength, we cannot exclude the possibility of the existence of multipoles exceeding the critical field strength. This inference aligns with our previous discussion on the observational examples.

\section{Phenomena Associated With Birth of Magnetar}\label{sec:grb}

Magnetars can theoretically form from binary NS mergers and from single star collapse. The formation of magnetar{s} is associated with and plays a critical role in various areas of astrophysics, including SLSN{e} \citep{2010ApJ...719L.204W,2010ApJ...717..245K}, GRBs \citep{2004RvMP...76.1143P,2014ARA&A..52...43B} and FRBs \citep{2022A&ARv..30....2P,2023RvMP...95c5005Z}. 

NS mergers (NS-NS) are expected to be important sources of gravitational waves \citep[GWs;][]{2017ApJ...851L..16A,2018A&A...615A..91B,2023PhRvX..13a1048A}, with the outcome depending on factors such as the total mass of the NS-NS system and the equation of state governing NSs. These mergers can lead to the formation of a black hole (BH), the creation of temporary hyper-massive and supra-massive NSs, or a stable NS \citep{2011ApJ...732L...6R,10.1093/mnras/stt2502,2014MNRAS.441.2433R,2016PhRvD..93d4065G}. For the single massive star, once the star depletes its nuclear fuel, the loss of pressure balance causes the core to collapse under its own gravity, then a NS or a magnetar could be born at the center \citep{1990RvMP...62..801B,2015SSRv..191..315M}. 

During the binary merger process and the single star collapse process, various mechanisms could amplify the magnetic field \citep{2021Univ....7..351I}. For example, {the dynamo mechanism converts kinetic energy into magnetic energy. In neutron-rich matter that is highly conductive, rapid rotation and convective movements create electric currents. These currents then produce very strong magnetic fields, ranging from $\sim 10^{14}$ to $10^{15}$ Gauss \citep{1992ApJ...392L...9D,2020SciA....6.2732R}.}

In the context of GRBs, long-duration GRBs (LGRBs) are often associated with type IC supernovae \citep{2012grb..book..169H,2014A&A...569A.108K,2017AdAst2017E...5C,2023ApJ...955...93A,2019ApJ...874...39W}, leaving behind a newborn NS at the center of the explosion, while short-duration bursts (SGRBs) are hypothesized to originate from mergers of compact object binaries (NS-NS/BH mergers)\citep[see e.g.,][and references therein]{2015PhR...561....1K,2018pgrb.book.....Z}.

Approximately half of the {{GRB} X-ray afterglows} exhibit the ``canonical'' pattern, characterized by a sequence of steep, shallow (plateau), and normal decay phases \citep{2006ApJ...642..389N,2006ApJ...642..354Z}. Moreover, a substantial majority of their X-ray emissions deviate{s} from the expected power-law decay \citep{1993ApJ...405..278M, 2009ApJ...707..328L, 2014A&A...565A..72M, Li2015, BERNARDINI201564}, showing flaring activities during the afterglow \citep{10.1111/j.1365-2966.2010.17037.x, 2018ApJ...852...53R}.

These observations pose challenges within the framework of a black hole central engine but are consistent with the concept of a rapidly spinning millisecond magnetar as the central engine \citep{2008MNRAS.385.1455M, 2010MNRAS.409..531R}. The magnetar central engine hypothesis offers a plausible explanation for the observed X-ray plateau during the GRB afterglow, as the magnetar is anticipated to dissipate its rotational energy \citep{1992Natur.357..472U,2011MNRAS.413.2031M, 2018pgrb.book.....Z,2022ApJ...936..190W,2023ApJ...945...95W}.

Furthermore, ``internal plateau'', {a phase at the beginning of the X-ray afterglow  where the flux remains nearly constant or decays very slowly then followed by a rapid drop, has been observed} in specific LGRBs and SGRBs.  \citep{2007ApJ...665..599T, 2013MNRAS.430.1061R, 2019ApJS..245....1T, 2020ApJ...901...75D, 2018qcs..confa1025L}.

{The decay slope following the “internal plateau” is too steep to be explained by the external shock model. This indicates that the plateau likely has an “internal” origin, implying that the central engine remains active for hundreds of seconds or sometimes longer before suddenly shutting down. Based on this logic, both the black hole and magnetar central engine models can provide many corresponding explanations.}

{If the central engine is a black hole, one proposal is that initially the fallback accretion rate is sufficiently high to maintain the jet power. However, as the accretion rate decreases over time, it eventually reaches a critical threshold beyond which the accretion can no longer confine the magnetic flux on the black hole. Consequently, the magnetic flux diffuses outwards, leading to a rapid decline in jet power and thus a sudden decay in the GRB luminosity \citep{2015MNRAS.447..327T, 2015ApJ...804L..16K, 2024ApJ...964..169Z}. }

{Fallback accretion can also produce similar light curve evolution in the magnetar model. For instance, in addition to the dipole emission, \citet{2017MNRAS.470.4925G} considered the propeller mechanism, which involves the interaction between the magnetar’s magnetic field and the accretion disc. The corresponding propeller luminosity is calculated based on the energy lost by the magnetar in transferring angular momentum to the ejected disc material. The bright early-time emission is sustained by the high propeller luminosity, and producing a very steep decay is favored by optimizing fallback parameters to ensure a large, rapid initial accretion, maximizing propeller efficiency, using a high beaming fraction, and choosing a very rapid initial spin period. }

{Another feasible interpretation within the magnetar model} is that the NS is ``supra-massive'' at birth and subsequently undergoes collapse to a black hole after significant spin-down \citep{2015ApJ...805...89L,2018pgrb.book.....Z, 2021ApJ...922..102H}. {However, the issue of the lifespan of such a supramassive NS is persistent and lacks a straightforward solution \citep{2014MNRAS.441.2433R}. \citet{2021ApJ...920..109B} simulated the evolution of binary neutron star merger and compared with the obseravtion. They concludes that while supramassive neutron stars formed after binary neutron star mergers are theoretically plausible, they are unlikely to be the primary explanation for the observed internal X-ray plateaus in GRBs. The narrow duration distribution of internal plateaus around 10$^3$ seconds from the observation contradicts the expected wide range of survival times over approximately five orders of magnitude. Moreover, the absence of extremely bright radio transients, originating from the immense energy release of collapsing supramassive neutron stars have never been detected. We also note, GRB 070110, where the internal plateau was observed and initially interpreted as resulting from the collapse of a supermassive NS \citep{2017ApJ...849..119C}. However, \citet{2007ApJ...665..599T} found a continuous optical lightcurve without the rapid drop characteristic at the time of the X-ray internal plateau. This contradicts the assumption of the collapse of a supermassive NS because the entire change of the central engine are likely to cause evolutionary changes across all wavelengths. In short, the possibility that an unstable supermassive neutron star could form in certain binary mergers and impact the evolution of GRBs cannot be completely ruled out. However, it is almost certain that not all internal plateaus are caused by the collapse of a supermassive neutron star. Examining this topic in depth is beyond the scope of this paper and will be addressed in future studies.}

It should be noted that, prevailing in the above-mentioned works, following the classic formulation of \cite{1983bhwd.book.....S}, \citep[see also][]{1969Natur.221..454G, 1982RvMP...54....1M,1990ApJ...359..444F}, strong consideration has been given solely to the dipole component, with various higher order magnetic components being neglected.

\subsection{X-ray afterglow of GRBs observed by Swift}

Exploring the Swift database allows for the identification of additional GRBs exhibiting similar temporal characteristics, which could lead to a deeper understanding of the mechanisms driving these sources \cite[see e.g.,][]{10.1093/mnras/stt872,2017A&A...600A..98D,2020ApJ...893..148R}. 

The UK Swift Science Data Centre\footnote{\href{https://www.swift.ac.uk/xrt_products/index.php}{$\rm https://www.swift.ac.uk/xrt\_products/index.php$}}, which provides ongoing updates to the dataset initially presented by \cite{2009MNRAS.397.1177E} offers a comprehensive database containing various properties of GRBs, such as their spectrum, position, light curve, and parameters derived from the best temporal power-law fits .

We explored the list and targeted the long or short GRBs whose afterglows exhibit a distinctive plateau phase followed by a temporal decay, described by a power-law function of the form $\propto t^{\alpha}$, where $\alpha$ represents the power-law index. Up to March 2021, we have identified a total of 204 GRBs exhibiting an observed plateau structure in their afterglows.

\begin{figure}
\centering
\includegraphics[width=1\hsize,clip,angle=0]{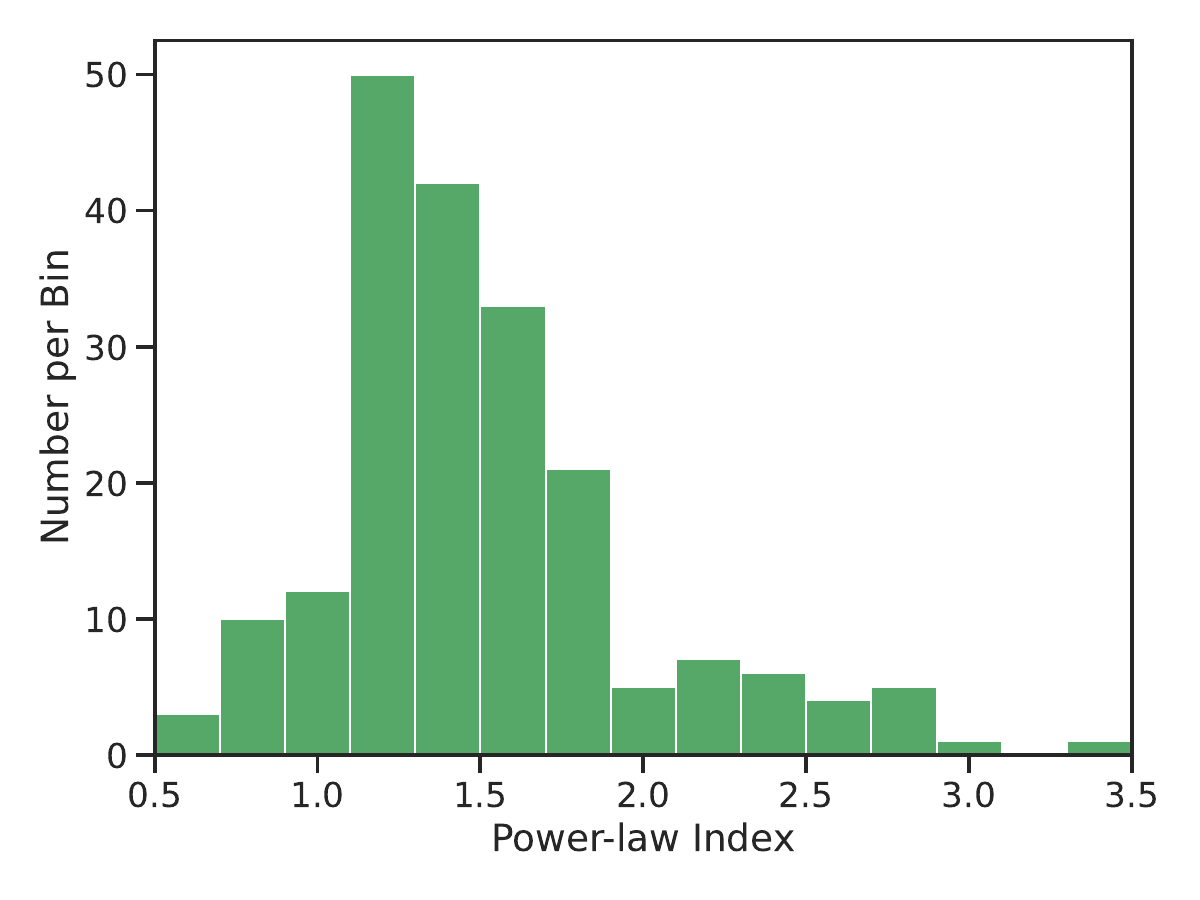}
\caption{The histogram of  power-law decay index of afterglow observed by Swift-XRT. In total 204 GRBs having a observed plateau structure till the end of March 2021. Most of the indices are between $-1$ and $-2$, the median and mean value are $-1.40$ and $-1.55$ respectively ({t}he index is defined as a negative value in this plot). }
\label{fig:plateau_hist}
\end{figure}

Figure~\ref{fig:plateau_hist} illustrates the distribution of power-law decay indices observed in the afterglows detected by Swift/XRT. The histogram indicates that the majority of the indices fall within the range of -1 to -2, with median and mean values of -1.40 and -1.55, respectively. This represents a common trend in the temporal decay behavior of these GRB afterglows, similar to the one predicted by higher orders of multipolar magnetic fields as discussed in the previous section. In the following section, we will investigate how this behavior can be effectively elucidated by the multipolar magnetic emissions.

\section{Fitting GRB Afterglow by Magnetar Multipolar Emission}\label{sec:application}

\begin{figure*}
\centering
\includegraphics[width=0.49\hsize,clip,angle=0]{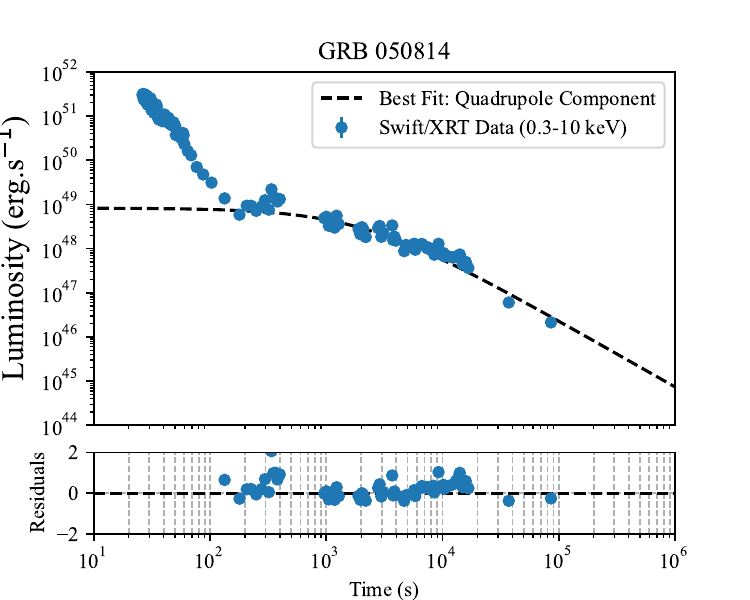}
\includegraphics[width=0.49\hsize,clip,angle=0]{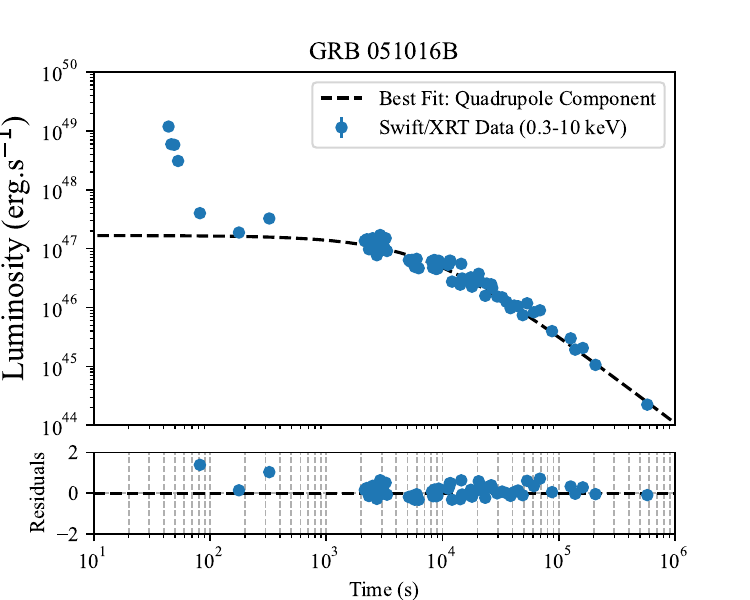}
\includegraphics[width=0.49\hsize,clip,angle=0]{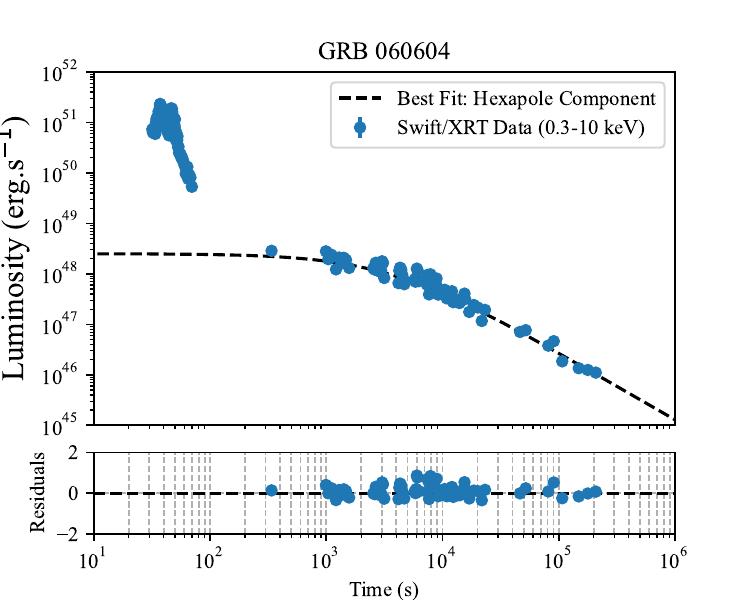}
\includegraphics[width=0.49\hsize,clip,angle=0]{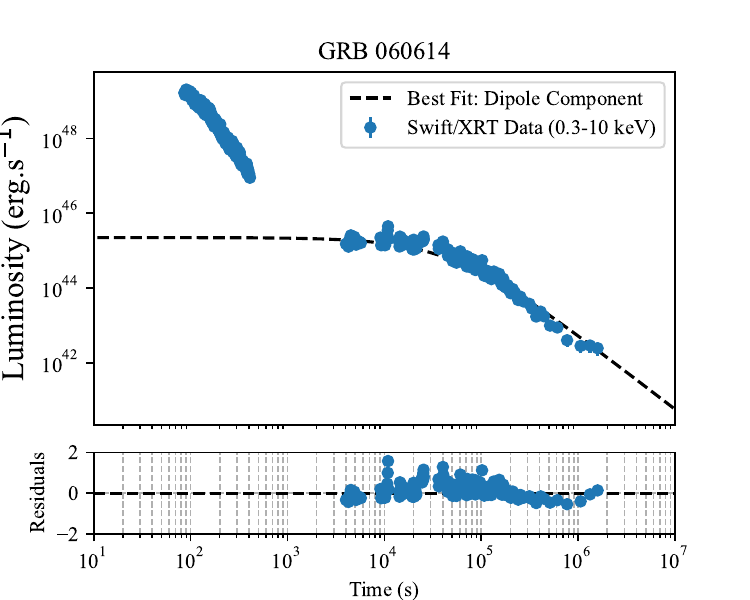}
\caption{ NS spin-down energy as the major energy source of the GRB {afterglow}. The multipolar model surpasses the traditional dipole model in fitting the normal decay, {as the higher-order multipolar components} can inject energy (spin-down luminosity) following a wider range of power-law decay indices (-2 to -1). This figure shows examples with fitted initial spin period (P$_0$) and magnetic field components {as reported in Table~\ref{tab:1}. In GRB 050814 and GRB 051016B, the quadrupole component is dominant; in GRB 060614, the hexapole component is dominant; and in GRB 060614, the dipole component is dominant. Since the plots are in logarithmic scale, the residuals are expressed relative to the model-predicted luminosity. Specifically, the residuals are defined as:
$\text{Residual} = \frac{\text{Observed Luminosity} - \text{Model-Predicted Luminosity}}{\text{Model-Predicted Luminosity}}$. Three out of four samples} suggest that fast-spinning newborn magnetars evolve high-order {multi}poles.}
\label{fig:multipole-luminosity-fitting}
\end{figure*}

\begin{deluxetable*}{l|cccccc}
\tablecaption{{Magnetic field components (B), chi-squared per degree of freedom ($\chi^2/\text{dof}$) of the best fit, and initial period (P$_0$) for GRB 050814, GRB 051016B, GRB 060604, and GRB 060614}\label{tab:1}}
\tablewidth{100pt}
\tablehead{
\colhead{GRB Name} \vline & \colhead{B$_{\text{Dipole}}$ (G)} & \colhead{B$_{\text{quad}}$ (G)} & \colhead{B$_{\text{hexa}}$ (G)} & 
\colhead{B$_{\text{octo}}$ (G)} & \colhead{$\chi^2/\text{dof}$} &
\colhead{P$_0$ (ms)}
}
\startdata
GRB 050814 & $-$ & $1.12 \times 10^{15}$ & $-$ & $-$ & 107/57 & 0.82 \\
GRB 051016B & $-$ & $4.25 \times 10^{15}$ & $-$ & $-$ & 72/63 & 2.9 \\
GRB 060604 & $-$ & $-$ & $1.36 \times 10^{15}$ & $-$ & 69/68 & 0.93 \\
GRB 060614 & $1.47 \times 10^{15}$ & $-$ & $-$ & $-$ & 277/153 & 13.8 \\
\enddata
\end{deluxetable*}

In the {discussion} presented in section \ref{sec:evolution}, it was highlighted that for magnetars with millisecond spins, the luminosity from multipole spindown can match or surpass that from dipole emission, and the lightcurve is composed of plateau followed by power-law decay, the decay index ranges between $-1$ and $-2$. The discussion in section \ref{sec:grb} explains that GRB{s} can give birth to magnetar{s} and shows consistent pattern of plateau plus decay observed in the GRB afterglow, noting that the decay index of GRB afterglows again falls within the $-1$ to $-2$ range. Additionally, the light curve of CDF-S XT2 \citep{2019Natur.568..198X}, the uniquely identified early-stage magnetar, exemplifies typical GRB afterglow features both temporally and energetically. In fact, CDF-S XT2 is believed to be the product of a binary NS merger. It probably has also produced a GRB, but the GRB is missed because the relativistic jet did not point to the satellites.

The above evidences prompt us to speculate that the plateau and subsequent normal decay phases of GRB afterglow are originated from the multipolar spindown energy of newborn magnetar. 
%,2010ApJ...718L..14S,2015ApJ...798...10R,2021ApJ...911...76X
We investigate this assumption for several  GRBs of well observed X-ray afterglow data by Swift, including {GRB 050814, GRB 051016B, GRB 060604, and} GRB 060614. \citep{2019ApJS..245....1T, 2015NatCo...6.7323Y}. 

We do not consider the radiation from the first few hundred seconds as this early radiation, {with a very steep decay shape, is considered the tail of the prompt emission. \citep{2006MNRAS.369..311Y,2006ApJ...642..354Z}}. We posit that the ratio of energy input from {multipolar spin-down emission} to the observed energy in the afterglow is unity, due to the involvement of several uncertain parameters. Firstly, the radiation from the afterglow is beamed, but the angle of beaming cannot be precisely constrained. Ignoring the beaming effect leads to an increase in both the estimated spin period and the magnetic field strength \citep{2010MNRAS.409..531R}. These adjustments counterbalance the effect of not considering the conversion efficiency from rotational energy to radiative energy, which is also an uncertain value \citep{2021ApJ...909L...3Z,2023ApJ...944L..57L}. This simplified treatment may bring an uncertainty of a factor of 3 \citep{2013MNRAS.430.1061R}.

{The most common theoretical model for afterglow is the external shock accelerates electrons to produce synchrotron radiation \citep{1998ApJ...497L..17S,2002ApJ...566..712Z} in various energy bands, including radio, optical, and X-ray \citep{2012ApJ...746..156C,2018ApJS..234...26L,2007A&A...469..379E}. Therefore, we cannot fit only the X-ray data; instead, we need to fit the bolometric luminosity. However, due to observational limitations, bolometric luminosity is not a value that can be precisely obtained. Based on past experience, we know that the lower energy limit is in the radio band \citep{2003AJ....125.2299F}, while the upper limit, according to NuStar observations, exceeds 79 keV \citep{2013ApJ...779L...1K}. Combining this with the synchrotron radiation model \citep{1998ApJ...497L..17S}, we estimate that the bolometric luminosity is $\sim 5$ times the energy contained in the Swift-XRT band (0.3-10 keV). Given that Swift-XRT typically provides the most comprehensive observations of afterglow evolution, we use the luminosity afterglow observed by Swift-XRT multiplied by $5$ as the bolometric luminosity.}

In figure~\ref{fig:multipole-luminosity-fitting}, the {bolometric} luminosities of these {GRB} afterglows are depicted, along with the fitted curves by the multipolar emission. {For data fitting, we utilized LMFIT \citep{2014zndo.....11813N}, which is an interface for performing non-linear optimization and curve fitting in Python\footnote{\href{https://lmfit.github.io/lmfit-py/}{https://lmfit.github.io/lmfit-py/}}}. It can be seen that the model fits the observational data of the plateau and normal decay phases very well.  The fittings reveal that {GRB 050814 and GRB 051016B could be effectively modeled using a quadrupole emission, resulting in magnetar spins of $\sim 1$ ms and magnetic field strengths of $1.12 \times 10^{15}$ Gauss and $4.25 \times 10^{15}$ Gauss, respectively. GRB 060614 aligns with hexapole emission, showing a spin of $\sim 1 $ ms and a magnetic field strength of $1.36 \times 10^{15}$ Gauss. GRB 060614 fits a dipole model with a spin of $13.8$ ms and a magnetic field strength of $1.47 \times 10^{15}$ Gauss; see Fig.~\ref{fig:multipole-luminosity-fitting} and Table.~\ref{tab:1}.}

{For each GRB, we have searched among the different configurations of the magnetic field and performed statistical tests to find the best fit. We consider the purely dipolar magnetic field fit as the baseline model and add higher-order components to see if their inclusion improves the fit. Therefore, we use the statistical F-test for nested models: the simpler or "restricted" model fits the luminosity with only the purely dipolar field ($\rm L_{dip}$), while the more complex or ``unrestricted'' model includes the dipolar field plus higher-order magnetic fields ($\rm L_{dip}+L_{quad}+L_{hexa}$). If the p-value of the F-test for nested models is less than 0.05, we consider it the best model \citep{casella2002statistical}. }

{Moreover, when a higher-order magnetic field component is dominant in the statistical fitting, we perform the fit using only that component and confirm that it has a lower reduced chi-squared than the purely dipolar magnetic field fit. Table.~\ref{tab:1} shows the fit parameters for the best fit of the luminosity of each GRB. As an indicator for finding the best fit, we have provided the reduced chi-squared ($\chi^2_{\rm red}$) of the fits, which is defined as the chi-squared divided by the degrees of freedom ($\chi^2/\rm dof$). The free parameters of the fits include the strengths of the dipole, quadrupole, and hexapole magnetic fields, as well as the initial angular velocity or period. }

{The results presented in Table~\ref{tab:1} indicate that the fits for GRB 051016B and GRB 060604 are satisfactory, while those for GRB 050814 and GRB 060614 are less convincing, with a reduced chi-squared, $\chi^2_{\rm red}$, exceeding 1.8. In the residual plots shown in Fig.~\ref{fig:multipole-luminosity-fitting}, a notable deviation from the model is observed for GRB 050814 at approximately $2-5 \times 10^2$ seconds. This deviation can potentially be attributed to residual contributions from the early phase decay or an X-ray flare activity, which is distinct from the plateau and late decay phases \citep[see e.g.,][]{2021ApJ...906...60Z}. Upon excluding this X-ray flare from the fitting procedure, the reduced chi-squared improves to $\chi^2_{\rm red} = 1.44$.}

{For GRB 060614, the residuals exhibit a pronounced curved pattern around the model, indicating the potential presence of at least one additional component. This is primarily attributed to a late X-ray bump occurring around $10^6$ seconds. The origin of this X-ray bump is not apparent; however, it is noteworthy that its timing coincides with a late near-infrared bump, which has been associated with possible macronova activity in the late afterglow of the long-short burst GRB 060614 \citep{2015NatCo...6.7323Y}. Upon excluding this X-ray bump from the fitting process, the residuals become more evenly distributed around the best fit, and the reduced chi-squared decreases to $\chi^2_{\rm red} = 1.64$. }

The {initial spins inferred for GRB 050814, GRB 051016B, and GRB 060604} are all around $1$ ms, which may be related to their afterglows being among the brightest in all observed GRBs. The light curve of GRB 060614 resembles that of the magnetar CDF-S XT2 \citep{2019Natur.568..198X}. CDF-S XT2 is located at a redshift of z=0.738, on the outskirts of its host galaxy \citep{2017ApJ...849..127Z}. The light curve of CDF-S XT2 features a plateau of luminosity in the order of $10^{45} \text{erg s}^{-1}$ that persists for $\sim 2 \times 10^3$~s, succeeded by a sharp power-law decline of index $-2.16^{+0.25}_{-0.29}$ \citep{2019Natur.568..198X}. GRB 060614 exhibits very similar luminosity strength and evolutionary behavior, except it has a longer duration of the plateau. The steep decay of the initial few hundred seconds observed in GRB 060614 but missed in CDF-S XT2 is reasonable, since in the case of CDF-S XT2, the jet which produces the initial steep decay is off-axis. GRB 060614 lacks the supernova association {which is a signal of a collapsar} \citep{2006Natur.444.1053G} but it is associated with a kilonova {which comes from $r$-process by the collision of neutron-rich matter} \citep{2015NatCo...6.7323Y,2016EPJWC.10908002J}, suggesting it has the same origin as CDF-S XT2, that {it is resulted from the merger of binary neutron stars}. 

{To obtain the observational duration of the plateau and compare it with the duration obtained from Eq.~\ref{eq:ctime}, we follow the procedure introduced in \citep{2009MNRAS.397.1177E, 2012ApJ...758...27L, 2016ApJS..224...20Y, 2017A&A...600A..98D, 2019ApJS..245....1T}. This procedure fits the light curve of each GRB with a smoothly broken power-law function, where the characteristic bending time is considered the observed end time of the plateau phase. For example, the observed plateau time for GRB 060614 is T$_{\rm plateau} = 4.8 \times 10^{4}$ s \citep{2019ApJS..245....1T}, and the plateau time obtained from Eq.~\ref{eq:ctime} is T$_{\rm plateau} = 5.03 \times 10^{4}$ s, which is quite similar to the observed value.}

{Statistically speaking, our fitting shows the dominance of a single magnetic field component, as shown in Table~\ref{tab:1}. Upon careful examination, adding other components does not change the $\chi^2/\text{dof}$ in a statistically significant manner. Therefore, constraining the parameters of secondary components is not plausible based on Swift/XRT data, which typically spans from approximately $10^1$~s to $10^6$~s after the trigger time. For instance, the hexapolar component of $B_{\rm hexa} = (1.36\pm 0.12) \times 10^{15}$ Gauss, dominates GRB 060604, with $\chi^2/\text{dof} = 69.08/68$. Adding another component, such as the dipolar component with $B_{\rm dip} = (1.23\pm 0.31) \times 10^{14}$ Gauss, results in $\chi^2/\text{dof} = 69.52/67$, leading to a p-value of 0.5. This indicates that the inclusion of the dipolar field is not statistically significant.}

{It is worth noting that during the fitting procedure, the GRB trigger times are set according to Swift, and changing the trigger time by a few seconds does not result in statistically significant differences. For instance, when testing GRB 050814, which has the shortest plateau duration among the GRBs shown in Fig.~\ref{fig:multipole-luminosity-fitting}, adjusting the trigger time by $10^2$~s does not significantly alter the results. The quadrupolar magnetic field still provides the best fit, though with a different value and initial period, which in turn affects the duration of the plateau, as expected from Eq.~\ref{eq:ctime}.
}

%To conclude, the fitting parameters obtained for each GRB are consistent with the expected characteristics of newborn magnetars, lending support to the magnetar origin hypothesis for GRB afterglows. Furthermore, these results provide reciprocal support for the theories that GRBs can lead to the formation of magnetars.

%%%%%%%%%%%%%%%%%%%%%%%%%%%%%%%%%%%%%%%%%%%%%%%%%%%%%%%%%%%%%
\section{Discussion and Conclusions}\label{sec:conclusion}
%%%%%%%%%%%%%%%%%%%%%%%%%%%%%%%%%%%%%%%%%%%%%%%%%%%%%%%%%%%%%

Our article is dedicated to investigating the multipolar radiation features of newly formed millisecond magnetars, specifically within the first year and even the first day following their birth. As highlighted in the Introduction, the role of multipolar emission is becoming increasingly prominent in explaining the properties of fast rotating NSs and the radiation in related phenomena, driven by advances in theoretical and computer simulations, as well as rapid growth in observations \citep{1992Natur.357..472U,2005ApJ...624L.109Z,2017Sci...355..817I, 2019ApJ...887L..23B,2022ARA&A..60..495P,2023RvMP...95c5005Z}.

In Sections~\ref{sec:model} and \ref{sec:evolution}, we have presented an analytical solution for the evolution of NS multipolar electromagnetic fields and their associated expected luminosities for a general magnetic field and mass of the NS. { Studying the complete physical setting of the problem can make the formulation quite complex. To capture the basic idea and simplify the model, this paper relies on a series of simplifications.}

{ Specifically, the solution is assumed to be in a vacuum. We have employed a simplified approach by treating the NS as a point-like source, rather than considering its finite size. Additionally, we have neglected the temporal evolution of the magnetic field and consistently ignored its toroidal component. We also assume that the energy released in the afterglow is entirely supplied by multipolar spin-down emission, and beam effects are disregarded. Given that the observational data do not extend far beyond $\rm 10^6$ seconds, the observations are primarily fitted by a single multipolar component that dominates the magnetic field topology.} 

{For example, in a more complex model, the magnetic field decays and a toroidal component exists due to effects such as Ohmic dissipation, Hall drift, and ambipolar diffusion, which in turn leads to a reduction in the magnetic field strength and its associated multipolar radiation \citep[see e.g.,][and references therein]{1992ApJ...395..250G, 2006RPPh...69.2631H, 2024MNRAS.528.5178S}. Furthermore, the presence of plasma in the pulsar's magnetosphere can modify the magnetic field structure. \cite{2006ApJ...648L..51S} demonstrated through numerical solutions of the force-free relativistic magnetohydrodynamics equations in pulsar magnetospheres that the initial dipolar configuration can be altered. As a result, the observed magnetic field strength is approximately 1.5 times weaker than in the vacuum case. Additionally, the spin-down luminosity can lose up to twice as much power compared to the dipolar configuration, although it still adheres to the same functional form as the dipolar spin-down luminosity.}

As an indicative application we have demonstrated that the emission from millisecond magnetars can serve as a source of GRB afterglow, explaining the occurrence of the GRB X-ray plateau and its subsequent temporal decay with a power-law behavior. As shown in Section~\ref{sec:grb}, the majority of the power-law decay indices observed in the afterglows of GRBs detected by Swift/XRT fall within the range of -1 to -2, with median and mean values of -1.40 and -1.55, respectively, which aligns well with the higher orders of multipolar emission from a millisecond magnetar. {Some GRBs exhibit both internal plateau and normal plateau. If we explain the internal plateau using the assumption of collapse of a supermassive neutron star into a black hole, we will face the situation where there is no longer a magnetar to apply our model to explain the normal plateau. We prefer to consider that, at least for these GRBs, it is not necessary to assume an unstable supermassive neutron star. Instead, a neutron star produced by binary merger with a mass up to $\sim 2.2 M_{\odot}$
can still be stable \citep{2016PhR...621..127L,2017ApJ...850L..19M,2018ApJ...852L..25R}. The key condition for producing the internal plateau is that the central engine ceases activity or its activity no longer influences the radiation. This can be achieved through many mechanisms discussed in Section \ref{sec:grb} and the references therein, such as fallback accretion reaching a critical threshold.}

The complex behavior observed in the afterglow emission of GRBs, which significantly deviates from the simple power-law decay predicted by the standard afterglow model \citep{1993ApJ...405..278M}, has been elucidated. This highlights the potential of multipolar electromagnetic emissions from highly magnetized, rapidly rotating magnetars, in providing a more comprehensive understanding of the temporal characteristics of GRB afterglows and similar phenomena. {In Section \ref{sec:application}, we selected four GRBs with excellent observational data and fitted them using our model. The results indicate that these GRBs can be well-explained by the magnetar model with a millisecond spin and  a  dipole or multipole magnetic field of $\sim10^{15}$ Gauss. We also discussed the uncertainties of the fitting. Firstly, these uncertainties stem from the data, as we do not have perfect observations of the bolometric luminosity. Secondly, there are uncertainties related to the beaming and conversion efficiency of the radiation. These factors contribute to an overall uncertainty of a factor of $\sim3$. Despite the lack of precise values, our results are consistent in magnitude with the expectations for newly born magnetars and align with previous observations.} 

Our study is distinctive in that it diverges from the typical observation of approximately 1-second magnetars, in which the dipole component predominantly governs the spin down. Additionally, for a newly formed magnetar with a spin period of approximately 1 millisecond, the multipolar luminosity is comparable to the dipole and significantly contributes to the spin down and subsequent afterglow radiation of GRBs. The high luminosity from the multipole components, coupled with their flexible decaying behavior, allows us to effectively model various GRB afterglows, including the plateau and subsequent normal decay phases.

In principle, the prominence of higher orders of multipole radiation in highly rotating NSs, specifically those with periods P $\lesssim  1~\text{ms}$, introduces additional important considerations. In the case of slowly rotating NSs, the moment of inertia remains constant. However, for rapidly rotating NSs, their oblateness causes significant variations in its moment of inertia as the period evolves, leading to notable observational implications \citep{1983bhwd.book.....S}. 

For example, it has been demonstrated by \cite{1990ApJ...359..444F} that during the initial stages of the evolution of a rapidly rotating star, the rotational frequency may increase even as the angular momentum decreases. This phenomenon is dependent on the specific nuclear equation of state adopted and could offer a potential explanation for the early flares superimposed on the afterglow emission observed in the X-rays of GRBs. However, a comprehensive investigation of this subject requires a separate study involving in-depth analytic and numerical treatments. Such an exploration could provide valuable insights into the underlying mechanisms driving the observed temporal behaviors in GRB afterglows and further enhance our understanding of the complex dynamics of highly rotating NSs.

\acknowledgments

{We would like to thank the anonymous referee for his valuable comments, which have significantly enhanced the quality of the paper, improved its clarity, and strengthened its technical content.} R. Moradi acknowledges support from the Institute of High Energy Physics, Chinese Academy of Sciences (E32984U810).

\bibliographystyle{aasjournal}
%\bibliography{multipole}

\end{document}